\documentclass[12pt,a4paper,english]{article}
\usepackage{babel,cite} \usepackage[dvips]{epsfig}

\makeatletter

\usepackage{amssymb,amsthm}

\newcommand{\be}{\begin{equation}} 
\newcommand{\ee}{\end{equation}}
\newcommand{\eq}[1]{(\ref{#1})}

\def\nn{\nonumber} 
\def\bea{\begin{eqnarray}} 
\def\eea{\end{eqnarray}}
\newcommand{\barr}{\begin{array}} 
\newcommand{\earr}{\end{array}}
 
\def\one{\mbox{1 \kern-.59em {\rm l}}}
\def\und#1{\underline{#1}}

\def\({\left(} 
\def\){\right)} 
\def\[{\left[} 
\def\]{\right]}

\def\a{\alpha}  
  
 \def\d{\delta}

  \def\la{\lambda}

\def\R{{\mathbb R}}  
 \def\one{\mbox{1 \kern-.59em {\rm l}}}

\def\cH{{\cal H}}      
    \def\cS{{\cal S}}  \def\cU{{\cal U}}

\def\bit{\begin{itemize}} \def\eit{\end{itemize}}

\def\diag{\mbox{diag}}

   \def\dd{\partial}

\makeatother
\begin{document}

\renewcommand{\title}[1]{\vspace{10mm}\noindent{\Large{\bf
#1}}\vspace{8mm}} \newcommand{\authors}[1]{\noindent{\large
#1}\vspace{5mm}} \newcommand{\address}[1]{{\itshape #1\vspace{2mm}}}


\begin{flushright}
UWThPh-2006-16
\end{flushright}

\begin{center}

\title{ \Large Exact renormalization of a noncommutative  \\[1ex]
$\phi^3$ model in 6 dimensions}

\vskip 3mm

\authors{Harald {\sc Grosse} and 
Harold {\sc Steinacker\footnote{Work supported by the FWF project P18657.}
}}

\vskip 3mm

\address{Institut f\"ur Theoretische Physik, Universit\"at Wien\\
Boltzmanngasse 5, A-1090 Wien, Austria}

\vskip 1.4cm

\textbf{Abstract}

\vskip 3mm

\begin{minipage}{14cm}

The noncommutative selfdual $\phi^3$ model in
6 dimensions is quantized and essentially solved, 
by mapping it to the Kontsevich model.  
The model is shown to be renormalizable and asymptotically free, 
and solvable genus by genus. 
It requires both wavefunction and coupling constant renormalization.
The exact (``all-order'') renormalization of the bare parameters is determined
explicitly, which turns out to depend on the genus 0 sector only. 
The running coupling constant is also computed exactly, 
which decreases more rapidly than 
predicted by the one-loop beta function. A phase transition to an
unstable phase is found.

\end{minipage}

\end{center}



\section{Introduction}

This paper is the third part of a series of papers 
\cite{Grosse:2005ig,Grosse:2006qv}
studying the quantization of a noncommutative ``self-dual'' $\phi^3$
model, by mapping it to the Kontsevich model. 
The model is characterized by an additional potential term
in the action, which takes care of the UV/IR mixing 
following
\cite{Grosse:2003nw,Grosse:2004yu,Grosse:2005da,Langmann:2002cc}. 
In the previous papers
we discussed the cases of 2 and 4 dimensions, and showed that the
model is renormalizable and essentially solvable. These cases are 
in some sense simpler because they are 
super-renormalizable. This is no longer the case in 6 dimensions, 
where new renormalization is required at each order, and the full
complexity of an interacting quantum field theory is found 
with both wavefunction and coupling constant renormalization.
Indeed, the commutative $\phi^3$ model in 6 dimensions is 
known \cite{Collins:1984xc} to be asymptotically free.

Generalizing \cite{Grosse:2005ig,Grosse:2006qv}, 
we show in this paper that the selfdual 
NC $\phi^3$ model in 6 dimensions 
can be renormalized and essentially solved 
in terms of a genus expansion. This is possible again
using the results of \cite{Itzykson:1992ya,Kontsevich:1992}
on the  Kontsevich model, 
which most however be properly renormalized. 
The model has 6 relevant parameters.
In particular, both coupling constant and wavefunction
renormalization are required in the 6-dimensional case,
in addition to tadpole and mass renormalization which were
sufficient in 2 and 4 dimensions. 
Remarkably, we find again that the renormalization is 
determined by the genus 0 sector only, 
as in the case of 2 and 4 dimensions. 

After renormalization, 
all $n$-point functions can be computed in principle in terms of
a genus 
expansion, and we give explicit expressions for the 1-, 2-, and some 
3-point functions. We also find a
 critical surface $\a=0$ in moduli space, which separates  
2 different phases.
This provides a model which contains the full
complexity of renormalization of a  
non-super-renormalizable asymptotically free 
quantum field theory, while 
being solvable and hence fully under control.
It is very remarkable that a 
6-dimensional interacting NC field theory 
allows such a detailed analytical description. 
It can therefore
also serve as a testing ground for various approximation methods,
which is of interest also in a more general context.

In particular, we are able to determine exactly the RG flow
of the bare parameters, as well as the running of the ``physical'' 
coupling constant i.e. the 1PI 3-point function.
We find that the 1-loop beta function 
for the coupling constant 
correctly predicts asymptotic
freedom, but wrongly predicts 
a $(\log N)^{-1/2}$ dependence on the scale
as opposed to the correct $(\log N)^{-2}$ dependence.

Beyond the selfdual case $\Omega =1$, 
perturbative renormalizability of similar models
has been established 
using variants of a renormalization group approach, see
e.g. \cite{Grosse:2003nw,Grosse:2004yu,Grosse:2005da,Gurau:2005gd,
Vignes-Tourneret:2006nb,Rivasseau:2005bh}. The present approach
using matrix model methods is certainly appropriate for the 
case $\Omega=1$. We indicate in section \ref{sec:beyondomega} how it might 
be extended to $\Omega\neq 1$, which however is not carried out here.
For a related matrix approach to NC field theory see also 
\cite{Steinacker:2005wj,Steinacker:2005tf}.

Being a continuation of  our previous work
on the 2- and 4-dimensional case,  we will be brief
in certain issues which have already been discussed there, and
which apply without change.
Nevertheless, the present paper is essentially self-contained. 
In section \ref{sec:phi3} we define the $\phi^3$ model
under consideration, and rewrite it as 
Kontsevich model. 
 We then briefly recall the most important facts about the
Kontsevich model in section \ref{sec:kontsevich}.
The main technical analysis is contained in  section
\ref{sec:renormaliz}, while the main results of this paper
are collected in section  \ref{sec:main}.
The asymptotic behavior of the 1-, 2- and 3-point functions 
is determined in section \ref{sec:asymptotic}, 
and further aspects such as relation with string field theory
are briefly discussed in  section  \ref{sec:further},
The perturbative results such as the one-loop beta-function
are given in section  \ref{sec:perturbative},
and we conclude with a general discussion and outlook.

\section{The noncommutative selfdual $\phi^3$ model}
\label{sec:phi3}

We consider the noncommutative 
$\phi^3$ model on the $6$-dimensional quantum plane $\R^{6}_{\theta}$,
which is generated by self-adjoint
operators\footnote{We ignore operator-technical subtleties, 
since the model will be regularized using a cutoff $N$.} $x_i$
satisfying the canonical commutation relations
\be 
[x_i,x_j] = i \theta_{ij}, 
\label{CCR}
\ee 
for $i,j=1,...,6$. We also introduce 
\be 
\tilde x_i = \theta^{-1}_{ij} x_j, \qquad [\tilde x_i,\tilde
x_j] = i \theta^{-1}_{ji}
\ee 
assuming that $\theta_{ij}$ is nondegenerate.
The model to be studied is defined by the action 
\be 
\tilde S = \int_{\R^{6}_{\theta}} \frac 12 \partial_i\phi
\partial_i\phi + \frac {\mu^2}2 \phi^2 + \Omega^2 (\tilde x_i \phi)
(\tilde x_i \phi) + \frac{i\lambda_0}{3!}\;\phi^3 .
\ee 
An additional oscillator-type potential 
 $\Omega^2 (\tilde x_i \phi)(\tilde x_i \phi) $ is included following 
\cite{Grosse:2003nw,Grosse:2004yu,Grosse:2005da,Langmann:2002cc}, 
making the model covariant under 
Langmann-Szabo duality, and taking care of the UV/IR mixing. 
The dynamical object is the scalar field
$\phi = \phi^\dagger$, which is  a self-adjoint operator acting
on the representation space $\cH$ of the algebra \eq{CCR}.
The action is written with imaginary coupling $i\la_0$,
so that the quantization is well-defined for real $\la_0$; otherwise  
the action would be unbounded. We will see however that after
quantization, one can perform an analytic continuation to  
real $\la_0' = i\la_0$.

Noting that the $\partial_i$ are inner derivatives
$\partial_i f = -i[\tilde x_i,f]$, the action can be written as 
\be 
\tilde S = \int -(\tilde
x_i\phi\tilde x_i\phi - \tilde x_i \tilde x_i \phi\phi) + \Omega^2
\tilde x_i \phi \tilde x_i \phi + \frac {\mu^2}2 \phi^2 +
\frac{i\lambda_0}{3!}\;\phi^3 
\ee 
using the cyclic property of the
integral. 
For the ``self-dual'' point $\Omega =1$, this simplifies further to 
\be
\tilde S = \int (\tilde x_i \tilde x_i + \frac
{\mu^2}2) \phi^2 + \frac{i\lambda_0}{3!}\;\phi^3 \,.
\label{action-E} 
\ee
In order to quantize the theory,  we need to include a linear (tadpole)
counterterm $-Tr (i\la) A\, \phi$ to the action (the explicit factor
$i\la$ is inserted for convenience).  
Replacing the integral by $\int \to (2\pi \theta)^3 Tr$
and adding a constant term for convenience,
we are led to consider the action
\be 
S = \,Tr \Big( \frac 12 J
\phi^2 + \frac{i \la}{3!}\;\phi^3 - (i\la) A \phi 
- \frac 1{3(i\la)^2}
J^3 -  J A\Big).
\label{action-kontsevich}
\ee 
Here
$J = 2 Z (2\pi \theta)^3 (\sum_i \tilde x_i \tilde x_i + \frac
{\mu^2}2 )$
is essentially 
the Hamiltonian of a 3-dimensional quantum mechanical harmonic
oscillator. A wavefunction renormalization $Z$ has also been introduced, 
which for the cubic term is absorbed in the 
redefined  coupling constant $\la$, in order to simplify
the notation. The field $\phi$ will  be the renormalized, finite
physical  field. In the usual basis of
eigenstates, $J$ then diagonalizes as 
\be
J |n_1,n_2,n_3\rangle 
= 16\pi^3 \theta^2 \,Z\, (n_1\;+ n_2\; + n_3\,  
+ \frac {\mu^2\theta+3}{2})|n_1,n_2,n_3\rangle, 
\qquad n_i \in \{0,1,2,...  \} 
\label{J-explicit}
\ee
assuming that $\theta_{ij}$ has the canonical form 
$\theta_{12} = -\theta_{21}\,= \theta_{34}= -\theta_{43} 
= ... =: \theta$. 
To simplify the notation, we will use the convention
\be
n \equiv (n_1,n_2,n_3), \qquad \und{n} \equiv n_1+n_2+n_3
\label{n-notation}
\ee
throughout this paper, keeping in mind that $n$ denotes  a
triple-index. It turns out that we need
\be
A = a_0 + a_1 J + a_2 J^2 .
\label{A-def}
\ee
By suitable shifts\footnote{for the
  quantization, the integral
for the diagonal elements is then defined via analytical continuation,
and the off-diagonal elements remain hermitian since $J$ is diagonal.}
 in the field $\phi$
\be 
\tilde\phi = \phi + \frac 1{i\la} J \, = \, X + \frac 1{i\la} M 
\label{phi-shift}
\ee 
one can now either eliminate
the linear term or the quadratic term in the action,
\be 
S= Tr \Big( -\frac 1{2 i \la} M^2 \tilde\phi + \frac{i
\la}{3!}\;\tilde\phi^3 \Big) 
= \,Tr \Big(\frac 12 M X^2 + \frac{i\la}{3!}\;X^3 -
\frac 1{3(i\la)^2} M^3 \Big)
\label{action-kontsevich-new}
\ee
where
\bea
M &=& \sqrt{J^2 + 2 (i\la)^2 A} 
= x Z\sqrt{\tilde J^2 + d} \label{M-def}\\ 
\tilde J &=& \frac{J}{Z} +
\frac{(i\lambda)^2\, a_1}{x^2 Z}, \label{undJ-def}\\ 
x &=& \sqrt{1+2(i\lambda)^2a_2},\label{x-def}\\ 
d &=& -\Big(\frac{(i\lambda)^2\,a_1}{x^2
Z}\Big)^2 + 2 \frac{(i\lambda)^2\, a_0}{x^2 Z^2 }.
\label{d-def} 
\eea
\eq{action-kontsevich-new} has precisely the form of the Kontsevich model
\cite{Kontsevich:1992}. The linear coupling of the field to the
source $M^2$ resp. $\tilde J^2$
will be very useful for computing correlation functions.
$\tilde J$ now has eigenvalues 
\be
\tilde J |n_1,n_2,n_3\rangle = 16\pi^3 \theta^2 \, (n_1\;+ n_2\; + n_3\,
+ \frac {\mu^2_R\theta+3}{2})|n_1,n_2,n_3\rangle
\label{und-J-explicit}
\ee 
which will be finite after renormalization, and 
\be 
\delta \mu^2
\theta = \mu^2\theta -\mu^2_R \theta = -\frac{2}{ 16\pi^3
\theta^2}\,\frac{(i\lambda)^2\,a_1}{x^2 Z} .
\label{delta-mu}
\ee

\subsection{Quantization, partition function  and correlators}
\label{sec:quantization}

The quantization of the model \eq{action-kontsevich}
resp. \eq{action-kontsevich-new} is defined by an integral over all
Hermitian $N^3\times N^3$ matrices $\phi$, where $N$ serves as a UV
and IR cutoff. The partition function 
\be Z(M) = \int
D\tilde \phi \,\exp\Big(- Tr \big( -\frac 1{2 i \la} M^2 \tilde\phi +
\frac{i\la}{3!}\;\tilde\phi^3 \big)\Big) = e^{F(M)}
\label{Z-again}
\ee 
defines the ``free energy'' $F(M)$, 
which is a function of the eigenvalues of $M$ resp. $\tilde J$.
Since $N$ is finite, we can freely switch between the various
parametrizations \eq{action-kontsevich}, \eq{action-kontsevich-new}
involving $M$, $J$, $\phi$, or $\tilde\phi$.  
Correlators or
``$n$-point functions'' are defined through 
\be \langle \phi_{i_1 j_1}
...  \phi_{i_n j_n}\rangle = \frac 1{Z(M)}\, \int D \phi \,\exp(- S)\,
\phi_{i_1 j_1} ....  \phi_{i_n j_n},
\label{correl-def}
\ee keeping in mind that each index denotes a multi-index
\eq{n-notation}.  Using the symmetry $Z(M) = Z(U^{-1} M U)$ for $U \in
U(N^3)$, we can assume that $M$ is diagonalized with ordered
eigenvalues $m_i$. There is a residual $U(1)\times U(3) \times U(6)
\times ...$ symmetry, reflecting the degeneracy of $J$. This implies
certain obvious ``index conservation laws'', such as $\langle\phi_{kl}
\rangle = \delta_{kl} \langle\phi_{ll} \rangle$ etc.

The nontrivial task is to show that  all these $n$-point functions with
finite indices have a well-defined and nontrivial limit $N \to
\infty$, for a suitable scaling of the bare
parameters. In addition, the index dependence of these $n$-point
functions must  be nontrivial.  
Recall that the wavefunction renormalization is already taken
into account in \eq{J-explicit}, so that $\phi$ is the finite, physical field.
The free parameters should in principle be determined by choosing
 renormalization conditions, such as $\langle\phi_{00} \phi_{00}\rangle = \frac
1{2\pi} \frac 1{\mu_R^2\theta +1}$, etc. These conditions can easily
be solved using the explicit results given in sections 
\ref{sec:renormaliz} and \ref{sec:main}, relating
the bare parameters with the ``physical'' quantities.


Noting that the field $\tilde \phi$ couples linearly to $M^2$
resp. $\tilde J^2$ in \eq{Z-again}, one can compute the connected
$n$-point functions by acting with the
derivative operator 
$2i\la \frac{\dd{}}{\dd J^2} = 2i\frac{\la}{x^2 Z^2} \frac{\dd{}}{\dd
  \tilde J^2}$ on $F$. Anticipating some results  below, 
we introduce the quantities
\be
i\la_R = \frac{i \la}{x^2 Z^{2}}, \qquad
r = \frac 1{x^2 Z}
\label{la-ren-def}
\ee
which turn out to be finite after renormalization. Then 
\be \langle
\tilde\phi_{i_1 j_1} ...  \tilde\phi_{i_n j_n}\rangle_c =
\big(2i\la_R\frac{\dd{}}{\dd \tilde J^2_{i_1 j_1} } \big)
.... \big( 2i\la_R\frac{\dd{}}{\dd \tilde J^2_{i_n j_n} }
\big) \,F(\tilde J^2).
\label{correl-gen-diag}
\ee 
Since the connected $n$-point functions are independent of the
shifts $\tilde \phi = \phi + \frac{J}{i\la}$ \eq{phi-shift} for $n
\geq 2$, the lhs coincides with the desired correlator $\langle
\phi_{i_1 j_1} ... \phi_{i_n j_n}\rangle_c$ for $n\geq 2$.
This strongly suggests that $\la_R$ should be finite.

Using the Dyson-Schwinger equations for the path integral \eq{Z-again}, one can
derive a number of nontrivial identities for the $n$-point
functions. Since their derivation was already given in
\cite{Grosse:2005ig}, we simply quote them here with the appropriate
substitutions.  In particular, one finds for the propagator \be
\langle\tilde \phi_{kl}\tilde \phi_{lk}\rangle = \frac
{2i\la}{m_k^2-m_l^2} \langle\tilde\phi_{kk}-\tilde \phi_{ll} \rangle
\label{Sdyson-3}
\ee for $k \neq l$ (no sum), where $m_k$ denotes the eigenvalues of
$M$. This gives 
\bea \langle\phi_{kl}\phi_{lk}\rangle &=&
\frac{2 r}{\tilde J_k + \tilde J_l} +
\frac{2i\la_R}{\tilde J_k^2-\tilde J_l^2}
\langle\phi_{kk}-\phi_{ll} \rangle
\label{2point-eom}
\eea 
using \eq{M-def}. The first term has the form of the free
contribution, and the
second is a quantum correction. The latter reduces to the 1-point 
functions, which will be obtained from the Kontsevich model.
Similarly, one can show \cite{Grosse:2005ig} 
\bea
\left<\tilde\phi_{kl}\tilde\phi_{lk} \tilde \phi_{kk}\right> &=&
\frac{2i\la}{m_k^2-m_l^2}\, \left<(\tilde \phi_{kk} - \tilde
\phi_{ll})\tilde \phi_{kk} - \tilde \phi_{kl} \tilde \phi_{lk}\right>
\label{3point-1}
\eea (no sum) for $k \neq l$. Using \eq{2point-eom}, this gives the
connected part \be \left<\phi_{kl}\phi_{lk} \phi_{kk}\right>_c =
\frac{2i\la}{m_k^2-m_l^2}\, \Big(\langle(\phi_{kk} - \phi_{ll})
\phi_{kk}\rangle_c - \langle\phi_{kl} \phi_{lk}\rangle\Big).
\label{3point-2}
\ee 
Clearly these relations can be generalized, greatly reducing the number of
independent correlators.  However, we will establish finiteness of the
general correlation functions directly, by showing that the
appropriate derivatives of the generating function
$F(\tilde J)$ are finite and well-defined after renormalization.

\section{The Kontsevich model: facts and background}
\label{sec:kontsevich}

The Kontsevich model is defined as a matrix integral
\be Z^{Kont}(\tilde
M)=e^{F^{Kont}}= \frac{\int dX \exp\left \{Tr \left(-\frac{\tilde M
X^2}{2} +i \frac{X^3}{6} \right)\right\} } {\int dX \exp\left\{-Tr
\left(\frac{\tilde M X^2}{2}\right)\right\} }
\label{Z-Konts}
\ee 
over hermitian $N^3 \times N^3$ matrices $X$, 
where the parameter $\tilde M$ is some given hermitian $N^3 \times N^3$ matrix.  This
model has been introduced by Kontsevich \cite{Kontsevich:1992} as a
combinatorial way of computing certain topological quantities
(intersection numbers) on moduli spaces of Riemann surfaces with
punctures.  It turns out to have an extremely rich structure related
to integrable models (KdV flows) and Virasoro constraints. For our
purpose, the most important results are those of
\cite{Kontsevich:1992,Makeenko:1991ec,Itzykson:1992ya} which provide
explicit expressions for the genus expansion of the free energy.  Note
that $\la$ can be introduced by rescaling the variables, writing $M =
\la^{2/3}\tilde M$.

The matrix integral in \eq{Z-Konts} and its large $N$ limit can be
defined rigorously in terms of its asymptotic series.  This involves
 the variables \cite{Kontsevich:1992} 
\be t_r :=
-(2r+1)!!\,\,\theta_{2r+1}, \qquad \theta_r := {1\over r} Tr \tilde
M^{-r}.  
\ee 
Then the large $N$ limit of the
partition function $Z^{Kont}(\tilde M)$ can be rigorously defined
\cite{Kontsevich:1992,Itzykson:1992ya}, which turns out to be a
function of these new variables only, $Z^{Kont}(\tilde M) =
Z^{Kont}(\theta_i)$. In our case, 
$\theta_r$ is divergent for $r=1,2,3$, which 
indicates that the model requires renormalization.

One can furthermore consider the genus expansion
 \be
\ln Z^{Kont} = F^{Kont} = \sum_{g\geq 0} F^{Kont}_g 
\ee 
by drawing the
Feynman diagrams on the appropriate Riemann surface, as usual for matrix
models.  
In \cite{Itzykson:1992ya} it was shown that all $F^{Kont}_g$ can be computed
using the KdV equations and the Virasoro constraints, which allows to 
find closed expressions for any genus $g$. They are given in terms of the
following variables 
\be 
I_k(u_0,t_i) = \sum_{p \geq 0} t_{k+p}
\frac{u_0^p}{p!}  = - (2k-1)!! \sum_{i} \frac 1{(\tilde m_i^2 - 2
u_0)^{k+\frac 12}},
\label{I-k-1}
\ee 
where $u_0$ is given by the solution of the implicit equation 
\be
u_0 = I_0(u_0,t_i) = -\sum_{i}
\frac{1}{\sqrt{\tilde m_i^2-2u_0}} 
\label{u-constraint}
\ee 
These variables turn out to be more useful for our purpose.  Using
the KdV equations, \cite{Itzykson:1992ya} obtain the following explicit
formulas: 
\bea F^{Kont}_0 &=& \frac{1}{3}\sum_{i} \tilde m_i^3 -
\frac{1}{3}\sum_{i} (\tilde m_i^2-2u_0)^{3/2} -u_0 \sum_{i}(\tilde
m_i^2-2u_0)^{1/2} \nonumber \\ &&+ \frac{u_0^3}{6}-\frac{1}{2}
\sum_{i,k}\ln\left\{\frac{(\tilde m_i^2-2u_0)^{1/2}+(\tilde
m_k^2-2u_0)^{1/2}} {\tilde m_i+\tilde m_k}\right\}
\label{F0Kont}\\
F^{Kont}_1 &=& \frac 1{24} \ln \frac 1{1-I_1}, \label{F-0-IZ-1}\\\
F^{Kont}_2 &=& \frac 1{5760}\left[5{I_4\over (1-I_1)^3} +29{I_3
I_2\over (1-I_1)^4 } +28{I_2^3\over (1-I_1)^5}\right]\ ,
\label{F-higher}
\eea 
etc. This form of $F_0$ was first found in \cite{Makeenko:1991ec}.
The sums over multi-indices \eq{n-notation} are to be interpreted as
$$ \sum\limits_i \equiv \sum\limits_{i_1,i_2,i_3=0}^{N-1}
$$ truncating the harmonic oscillators in \eq{J-explicit}.  
For our purpose, the most important
result is that all
$F^{Kont}_g$ with $g \geq 2$ are given by polynomials $F^{Kont}_g
= \sum \chi_{\{l_k\}}\, x_2^{l_2} ... x_{3g-2}^{l_{3g-2}}\,$ in $x_k =
I_k/ (1-I_1)^{{2k+1\over 3}}$ with the constraint 
\be 
\sum_{2 \leq k\leq 3g-2} (k-1) l_k = 3g-3,
\label{highergenus-constraint}
\ee 
where $l_k$ is the power of $x_k$, and
$\chi_{\{l_k\}}$ is some (rational) intersection number.

While many of these expressions are divergent as $N\to\infty$ for
$\tilde J$
given by \eq{und-J-explicit}, the physically relevant observables will
be convergent after renormalization.

\section{The Kontsevich model applied to the $\phi^3$ model}
\label{sec:appplic}

In order to apply the above results to the noncommutative $\phi^3$ model,
we need the following slightly modified version of the 
Kontsevich model, corresponding to the action 
 \eq{action-kontsevich-new}:
\be 
Z(M) = \exp(F(M)) = Z^{Kont}[\tilde M] Z^{free}[\tilde M]
\exp(\frac 1{3(i\la)^2} Tr M^3) \ee 
where 
\bea 
F_0 &:=& F_0^{Kont} +
F_{free} + \frac 1{3(i\la)^2} Tr M^3 \nn\\ 
&=& - \frac{1}{3}\sum_{i}
\sqrt{\tilde m_i^2-2u_0}^3 -u_0 \sum_{i}\sqrt{\tilde m_i^2-2u_0}
\nonumber \\ 
&&+ \frac{u_0^3}{6}-\frac{1}{2}
\sum_{i,k}\ln(\sqrt{\tilde m_i^2-2u_0}+\sqrt{\tilde m_k^2-2u_0}).
\label{F0}
\eea 
and $F_g = F_g^{Kont}$ for $g\geq 1$.
In the present case, the eigenvalues $\tilde m_i$ are given by
\eq{M-def} 
\be \tilde m_i  = \la^{-2/3} xZ \sqrt{\tilde J_i^2 + d} .
\ee 
The model as it stands is 
ill-defined for $N \to \infty$, since $u_0$ and 
many of the above sums are 
divergent. However, we can recast \eq{F0} using more appropriate 
variables, which suggests how to renormalize 
the various parameters, i.e. how to scale them with $N$.
Note that only the combinations $\sqrt{\tilde
m_i^2 -2u_0}$ enter in \eq{I-k-1} and \eq{F0}, which can be rewritten
using \eq{M-def} as 
\be 
\sqrt{\tilde m_i^2 -2u_0} = x Z
\la^{-2/3}\sqrt{\tilde J_k^2 + 2 b } .
\ee 
Here $b$ is defined
through 
\be \frac{x^3 Z^3}{(i\la)^2}\,\( b  - \frac 12 d\) = x
Z\,\la^{-2/3}\, u_0 = - \sum_{i} \frac 1{\sqrt{\tilde J_i^2 + 2 b }}
\label{b-def}
\ee 
using the constraint \eq{u-constraint}, which is replaced by
\eq{b-def} henceforth.  Eliminating $u_0$ using \eq{b-def} and
expressing $\la$ in terms of $\la_R$, $F_0$
takes the form 
\bea 
F_0 &=& \frac{1}{3(i\la_R)^{2}xZ}\sum_{i}
\sqrt{\tilde J_i^2+2 b }^3 -\frac{ b -\frac 12 d}{(i\lambda_R)^2 xZ} 
 \sum_{i}\sqrt{\tilde J_i^2 +2 b } \nn \\ 
&&
\!\!\!\!\!\!\!\!-\frac{( b -\frac 12 d)^3}{6(i\la_R)^{4} x^2 Z^2} 
 -\frac{1}{2} \sum_{i,k}\ln
\frac 1{(\la_R^{2}xZ)^{1/3}}\left(\sqrt{\tilde J_i^2+2 b } +\sqrt{\tilde
J_k^2+2 b }\right).
\label{F0tilde-2}
\eea 
The quantities $\tilde J, \, b $ and $\la_R$ 
will be finite after renormalization,
rendering the model well-defined.  We consider $F = F(\tilde J^2)$ as a
function of (the eigenvalues of) $\tilde J^2$ from now on, while $b$ 
is implicitly determined by \eq{b-def}.  
Since the eigenvalues $\tilde J_k$ only
enter through the combination $\sqrt{\tilde J_k^2 +2 b }$, we note
that the eigenvalues can be analytically continued as long as this
square-root is well-defined.

We can now compute various $n$-point functions by taking partial
derivatives of $F = \sum_g F_g$ 
with respect to $\tilde J^2$, as indicated in section
\ref{sec:quantization}. For the ``diagonal'' $n$-point functions
$\langle\tilde\phi_{ii} ... \tilde\phi_{kk}\rangle_c$, this amounts to
varying the eigenvalues $\tilde J_k^2$.  In doing so, we must
remember that $ b $ depends implicitly on $\tilde J_k^2$ through the
constraint \eq{b-def}.  However, some of these computations simplify
recalling that the constraint \eq{b-def} for $ b $ arises
automatically through the e.o.m: using 
\be \frac{\partial}{\partial
 b } \, F_0(\tilde J_i^2; b ) 
= -\frac 12 \Big(\frac{ b -\frac 12 d}{(i\la_R)^2 xZ} 
+ \sum_{i}\frac 1{\sqrt{\tilde J_i^2 + 2  b }}\,\Big)^2 =0
\label{constr-implicit-2}
\ee 
we can write \be \frac{d}{d \tilde J_i^2} F_0(\tilde J_i^2) =
\frac{\partial}{\partial \tilde J_i^2} F_0(\tilde J_i^2; b ) +
\frac{\partial}{\partial  b } F_0(\tilde J_i^2; b )
\frac{\partial}{\partial \tilde J_i^2} b  =\frac{\partial}{\partial
\tilde J_i^2} F_0( \tilde J_i^2; b ) .
\label{Fkont-partial}
\ee Thus for derivatives of order $\leq 2$ w.r.t. $ \tilde J_k^2$, we
can simply ignore $ b $ and treat it as independent variable, since
the omitted terms \eq{constr-implicit-2} vanish anyway once the
constraint is imposed.

\subsection{Renormalization}
\label{sec:renormaliz}

This section contains some computations required to show
finiteness of the $n$-point functions for any genus $g$.
This will establish Theorem 1 in section \ref{sec:main}.

\paragraph{The 1-point function}

We can now determine the required renormalization by considering the
one-point function. Using \eq{F0tilde-2}, \eq{constr-implicit-2} and
\eq{b-def}, the genus zero contribution is 
\bea 
\langle\phi_{kk}\rangle_{g=0} &=& 2i\la_R \frac{\dd{}}{\dd \tilde
J_k^2 } \,F_0(\tilde J^2) - \frac{J_k}{i\la_R x^2 Z^2} \nn\\ 
&=& \frac{1}{i\la_R x Z}\, y_k + (i\la_R)
\sum_{j}\frac{1}{y_k \sqrt{\tilde J_j^2+2 b } +
(\tilde J_j^2+2 b )} - \frac{r}{i\la_R}\,\tilde J_k+
i\lambda_R Z^2\,a_1 \nn\\
 &=:& W(y_k)
\label{F-derivative-2}
\eea 
using \eq{la-ren-def}, 
which must be finite and well-defined as $N \to
\infty$.  Here we define 
\be
y_k = \sqrt{\tilde J_k^2+2 b}.
\ee
We will find that $W(y)$ 
becomes a smooth function of $y$ after
renormalization, which amounts to the statement that the index
dependence of the one-point function is ``smooth''. This is a typical
feature of matrix models, reflecting some ``smooth'' distribution of
eigenvalues. This becomes here part of the statement of
renormalizability.
To proceed, we need to understand the function 
\be 
f(y) :=
\sum_{j}\frac{1}{y \sqrt{\tilde J_j^2+2 b } + (\tilde J_j^2+2 b )} 
\label{F-def}
\ee 
which as it stands is ill-defined for $N \to \infty$.

\subsubsection{Renormalization of  $f(y) $}
\label{sec:f-renormaliz}

In order to make sense of $f(y)$, we consider the
Taylor-expansion of 
\be f(y;\tilde J) = \sum_{j}\frac{1}{y \sqrt{\tilde
J_j^2+2 b } + (\tilde J_j^2+2 b )} = f_0(\tilde J) + y f_1(\tilde J) +
f_R(y; \tilde J)
\label{F-expansion}
\ee 
in $y$, in analogy to the usual strategy in renormalization. Here 
\bea 
f_0(\tilde J) &=& \sum_{j}\frac{1}{\tilde J_j^2+2 b } \, 
= f_0 + f_{0,R}(\d\tilde J) \label{f0-def}\\ 
f_1(\tilde J) &=& -\sum_{j}\frac{1}{(\tilde J_j^2+2 b )^{3/2}} \, = f_1 +
f_{1,R}(\d\tilde J) \label{f1-def}\\ 
f_R(y;\tilde J) &=& y^2 \sum_j \frac 1{(\tilde J_j^2+2 b )\Big(y
\sqrt{\tilde J_j^2+2 b } + (\tilde J_j^2+2 b )\Big)}. \label{fR-def}
\eea 
where  $f_1,f_2$ are divergent constants obtained by fixing $\tilde
J$ as in \eq{und-J-explicit}, while the
$f_{i,R}(\d\tilde J)$ are regular (convergent) 
functions obtained by taking into
account variations $\delta \tilde J$ of the eigenvalues. This is
necessary e.g. to compute partial derivatives w.r.t. $\tilde J$.
 Thus 
\bea
f_0 &=& \frac{1}{(16\pi^3 \theta^2)^2} \, \int_0^N dx_1 dx_2 dx_3\,
\frac 1{(x_1+x_2 +x_3 +\frac{3+\mu_R^2\theta}2)^2 } +finite \nn\\
&=& \frac{1}{(16\pi^3 \theta^2)^2} \, \Big((6\log(2)-3\log(3))N -
\frac{3+\mu_R^2\theta}2 \log(N) \Big) +finite 
\label{f0-scaling}
\eea 
and similarly
\bea f_1 = -\frac 12 \frac{1}{(16\pi^3 \theta^2)^3} \, \log(N) \,\, +
finite .  
\eea 
The remaining part $f_R(y;\tilde J)$ is well-defined and convergent,
provided $b$ and $\tilde J_k$ i.e. $\mu_R$ are finite, which will be 
assumed from now on\footnote{We do not pursue the possibility of 
divergent $\mu_R$, which does not appear to be interesting.}. 
To understand it better, we note that 
$$
f_R''(y) = f''(y) = 2\sum_{j}\frac{\tilde J_j^2+2 b }{\Big(y
\sqrt{\tilde J_j^2+2 b } + (\tilde J_j^2+2 b )\Big)^3} 
$$
is
positive, and similarly $f_R'(y) \geq 0$. 
Hence $f_R(y)$ is a rather simple
convex smooth function of $y>0$, which satisfies $f_R(0) = f_R'(0) =0$, 
and it remains only to determine
its asymptotic behavior for large $y \approx \tilde J_N$, i.e. small
$x$. This can be obtained by writing
\be
f_R(y) \approx
\frac{y}{(16\pi^3 \theta^2)^3} \, \int_{\tilde J_0/y}^{\tilde J_N/y} d^3 x
\frac 1{(x^2+2 b /y^2)\Big(\sqrt{x^2+2 b /y^2} + (x^2+2 b /y^2)\Big)}
\label{F-R-integral}
\ee
where $x = \frac{\tilde J}y$. This integral is convergent for
large $x$, but logarithmically divergent at the origin i.e. for large
$y \sim J_N$. We can hence replace the upper integration limit 
in \eq{F-R-integral} by $x=1$, 
to obtain the asymptotic behavior for large $y$:
\bea 
f_R(y) &\approx&
\frac{y}{(16\pi^3 \theta^2)^3} \, \int_{\tilde J_0/y}^{1} d^3 x \frac
1{(x^2+2 b /y^2)^{3/2}} 
\approx \frac 12
\frac{1}{(16\pi^3 \theta^2)^3} \,y \log(\frac y{\tilde J_0}).
\label{fR-asymptotic}
\eea

\subsubsection{Renormalization of  $\langle \phi_{kk}\rangle_{g=0} $}

We have seen that only $f_0$ and $f_1$ are divergent in
\eq{F-expansion}, while $f_R(y_k)$ is finite and well-defined as $N \to
\infty$.  Then \eq{F-derivative-2} becomes 
\bea 
\langle\phi_{kk}\rangle_{g=0} &=& \frac{y_k}{i\la_R x Z}  -
\frac{r \tilde J_k}{(i\la_R)} + i\lambda_R Z^2\,a_1 +
i\la_R ( f_0(\tilde J) + y_k\, f_1(\tilde J) +
f_R(y_k;\tilde J) ) \nn\\ 
&=& \(i\lambda_R Z^2\,a_1 +i\la_R f_0\) 
+ \Big(\frac{1}{i\la_R xZ} + i\la_R f_1 \Big)\,  y_k 
- \frac{r \tilde J_k}{i\la_R}  \nn\\ 
&& \quad + i\la_R \Big(f_R(y_k;\tilde J) + f_{0,R}(\d\tilde J) +
y_k f_{1,R}(\d\tilde J)\Big) \nn\\ 
&=& c -\frac{y_k}{ i\la_R \a}  - \frac{r \tilde J_k}{i\la_R}\, 
+ i\la_R \Big(f_R(y_k) + f_{0,R}(\d\tilde J) + y_k f_{1,R}(\d\tilde J)\Big)
\label{F-derivative-4}
\eea 
Here we define the quantities
\bea
\a &=&  -\frac{ x Z}{1 + (i\la_R)^2 x Z\, f_1} 
= -\frac{1}{\frac 1{xZ} + (i\la_R)^2 \, f_1} , \label{alpha-def}\\
c &=& i\la_R \, (Z^2\,a_1+ f_0).   \label{c-def}
\eea
Since $\tilde J_k, \, y_k$ and $f_R(y_k)$ are independent functions
of $k$ (as long as $ b  \neq 0$),  it follows that
$\langle\phi_{kk}\rangle_{g=0}$ is finite if and only if
the four quantities
\be
(i\la_R,\, c, \,\frac{1}{i\la_R \a},\, \frac{r}{i\la_R})
\label{onepoint-finite}
\ee
are finite.
Remarkably, the condition that $\a$ be finite will also 
guarantee that the higher genus contributions are finite.
The second form of \eq{alpha-def} implies that
this is possible only for real coupling $(i\la_R)$, 
since $x$ and $Z$ should be positive, while $f_1\sim - \log N$.

\subsubsection{Higher derivatives and higher genus contributions}

The connected part of the $n$-point 
functions for diagonal entries 
$\langle \tilde \phi_{i_1 i_1} ... \tilde \phi_{i_n i_n}\rangle_c$
are obtained by taking higher derivatives 
$(i\la_R)^n \frac{\dd{}}{\dd \tilde J^2_{i_1} } 
...  \frac{\dd{}}{\dd \tilde J^2_{i_n} }$
of $F(\tilde J^2)$ resp. \eq{F-derivative-4}.
Since the 
(infinite) shift $\tilde \phi = \phi + \frac{J}{i\la}$
drops out from the connected $n$-point function for $n \geq 2$, 
 these coincide with
$\langle \phi_{i_1 i_1} ... \phi_{i_n i_n}\rangle_c$ for $n \geq 2$,
and we have to show that they are finite. 

Consider first the genus 0 contributions. 
To compute higher derivatives of $F_0$ w.r.t. $\tilde J^2$, we 
must also take into account the implicit dependence of $b$ on $\tilde
J^2$. Indeed $b$ is a smooth function of $\tilde J$ as shown in 
\eq{b-derivative}. Furthermore,
recall that $\frac{\partial}{\partial  b } \, F_0(\tilde J_i^2; b )$
 vanishes through the constraint \eq{constr-implicit-2}, 
however this is no longer true for the higher derivatives. 
In particular, taking
 derivatives of \eq{constr-implicit-2} we find
\bea
-\frac{\partial^2}{\partial  b ^2} \, F_0(\tilde J_i^2; b ) &=&
\frac{\partial}{\partial  b  } \, 
\Big(\frac{ b -\frac 12 d}{(i\la_R)^2 xZ}
+ \sum_{i}\frac 1{\sqrt{\tilde J_i^2 + 2  b }}\,\Big)
 = \frac{1 - I_1}{(i\la_R)^2xZ}, \nn\\
-\frac{\partial^2}{\partial \tilde J_k^2 \partial  b } \, F_0(\tilde
J_i^2; b ) 
&=&\frac{\partial}{\partial \tilde J_k^2  } \,
\Big(\frac{ b -\frac 12 d}{(i\la_R)^2 xZ} 
+ \sum_{i}\frac 1{\sqrt{\tilde J_i^2 + 2  b }}\,\Big)
 =\frac 1{\sqrt{\tilde J_i^2 + 2  b }^3} 
\eea
which are both finite and smooth using \eq{1-I-reg-2} below. 
Combining this with
the explicit form \eq{F-derivative-4} 
and using the results of section \ref{sec:f-renormaliz}, it follows that
all higher derivatives of $F_0(\tilde J^2)$ 
w.r.t. $\tilde J^2_k $ are finite.

For the higher genus contributions,
we also need the derivatives of the quantities
\bea
I_k(\tilde J_i^2, b ) 
&=&   -(2k-1)!! \big(\la_R^{2} xZ\big)^{\frac{2k+1}3}
\sum_{i}\frac 1{(\tilde J_i^2 + 2 b )^{k+\frac 12}}.
\label{I-p-sum}
\eea
In particular, 
\bea
I_1(\tilde J_k^2, b ) &=&  -\la_R^{2} xZ
\sum_{i}\frac 1{(\tilde J_i^2 + 2 b )^{\frac 32}}\, 
= \la_R^{2} xZ \,f_1(\tilde J) \nn\\
&=&  \la_R^{2} xZ( f_1 + f_{1,R}(\d\tilde J))
\label{I-1-sum}
\eea
where $f_{1,R}(\d\tilde J)$ 
is a finite and smooth function \eq{f1-def} of $\tilde J_k^2$ and  $ b $
(which vanishes for $\delta\tilde J =0$).
Hence for $I_1$ to be finite we need
\be
\frac{1}{\la_R^{2} xZ} \sim f_1 \sim \log(N),
\label{la-scaling}
\ee
which using \eq{I-p-sum}
implies that all higher $I_k$ vanish for $N \to \infty$ unless
$(1-I_1) \sim 0$.
Hence the higher-genus contributions can be nontrivial only if
we carefully take a ``double-scaling'' limit\footnote{This is quite
  reminiscent of the double-scaling limit of matrix models in the
  context of 2D gravity} and require
$\frac{I_k}{(1-I_1)^{{2k+1\over 3}}} $
to be finite, while $I_1 \to 1$. 
Using \eq{I-1-sum}, we have
\be
\frac{1}{\la_R^{2} xZ} (1-I_1) 
= \frac{1}{\a(i\la_R)^2} - f_{1,R}(\d\tilde J)
\label{1-I-reg-2}
\ee
where $\a$ is defined in \eq{alpha-def}.
It follows that
\bea
\frac{I_k}{(1-I_1)^{{2k+1\over 3}}} 
&=&  - \frac {(2k-1)!!}
{\(\frac{1}{(i\la_R)^2 \a} -  f_{1,R}(\d\tilde J)\)^{\frac{2k+1}3}}\,
\sum_{i}\frac 1{(\tilde J_i^2 + 2 b )^{k+\frac 12}}  \nn\\
&\sim& - (2k-1)!!{\( (i\la_R)^2 \a\)^{\frac{2k+1}3}}\,
\sum_{i}\frac 1{(\tilde J_i^2 + 2 b )^{k+\frac 12}} 
\label{IkI1}
\eea
(the last form holds for $\delta\tilde J =0$)
is  finite and nontrivial for $k \geq 2$, provided
\be
\a = finite,
\ee
assuming that $i\la_R$ is finite.
In particular, all derivatives of $F_g$ for $g \geq 2$ w.r.t.
$\tilde J_k^2$ are manifestly finite, and thus all genus $g \geq 2$
contributions to the diagonal $n$-point functions are finite. 
Using \eq{1-I-reg-2} and \eq{F-0-IZ-1}, this also holds for genus 1.

Finally, from the constraint \eq{b-def} we derive
\bea
\frac{\partial}{\partial \tilde J_k^2}  b 
&=& -(i\la_R)^2 xZ\, \frac{\partial}{\partial \tilde J_k^2}
 \Big(\sum_{i} \frac 1{\sqrt{\tilde J_i^2 + 2 b }}\Big)  \nn\\
&=& \frac 12 (i\la_R)^2 xZ\, 
 \Big(\frac 1{\sqrt{\tilde J_k^2 + 2 b }^3} 
 + \sum_{i}\frac 2{\sqrt{\tilde J_k^2 + 2 b }^3}\,\frac{\partial}{\partial \tilde J_k^2}  b  \Big) 
\eea
which using \eq{1-I-reg-2} gives
\be
\frac{\partial}{\partial \tilde J_k^2}\,  b 
=  -\frac 12  
  \frac 1{\sqrt{\tilde J_k^2 + 2 b }^3}\,
\frac{1}{\frac{1}{(i\la_R)^2 \a} - f_{1,R}(\d\tilde J)}\,.
\label{b-derivative}
\ee
This is again finite, using  the above assumptions.
Hence $b$ depends smoothly on
the variations in $\tilde J_k$ provided $\a$ is finite. 
If $\a =\infty$,
then the constraint \eq{b-def} cannot be solved any more for $b$ as a function
of the eigenvalues $\tilde J$ (and their variation), rendering the 
model non-renormalizable. This implies that the genus 0
sector fully determines the required renormalization, as was found
previously in 2 and 4 dimensions \cite{Grosse:2006qv,Grosse:2005ig}.

\subsection{Main result and renormalization group flow}
\label{sec:main}

We have now established all required formulas for the derivatives 
of $F(\tilde J)$, and hence for the diagonal $n$-point functions.
We also showed that all of these are finite and have a 
well-defined as $N \to \infty$, provided a few renormalization conditions hold.
Let us collect these conditions and use them to
determine the required scaling of the bare parameters.

Assume first that $\la_R =0$. Imposing this exactly (i.e.
independent of $N$) implies $\a =-xZ$ and 
$i\la = i\la_R \a^2 =0$. This is the free case
with $x$ and $Z$ finite\footnote{We shall not pursue here the
  possibility 
of other more subtle scaling limits.},
as can be seen e.g. from \eq{F-derivative-4} and \eq{2point-exact}.

Hence assume $\la_R \neq 0$.
We assume also that $ b  \neq 0$ 
(we will see that this restriction is in fact not important), so that the 
4 quantities in \eq{onepoint-finite} are finite. By taking products, 
this implies  that 
$r$ and $\frac{1}{\a}$ are also finite, hence
$\a \neq 0$. Finite $\a$ is in fact also required for 
the higher-genus contributions to be finite.
Using \eq{alpha-def}, we then obtain 
\be
\frac{1}{x Z} = - (i\la_R)^2 f_1 \,  -\frac{1}{\a}
 =  \frac 12 \frac{(i\la_R)^2}{(16\pi^3 \theta^2)^3} \, \log(N)  \,
 -\frac{1}{\a}
\label{xZ-eq}
\ee
which implies\footnote{note also that \eq{xZ-eq} implies that $(i\la_R)$ should be real, 
since $xZ$ should be positive. }
\be
i \la = i\la_R x^2 Z^2 
= \frac{(i\la_R)}{\Big(-\frac{1}{\a} + \frac 12
  \frac{(i\la_R)^2\log(N)}{(16\pi^3 \theta^2)^3}\Big)^{2}} \quad
\sim (i\la_R)^{-3}\, \log(N)^{-2}
\label{la-flow}
\ee
for large $N$.
Note that in a perturbative approach i.e. 
if this is formally expanded in terms of $\la_R$, this is divergent and
requires renormalization at each order. Nevertheless the closed form 
\eq{la-flow} is rather simple, with leading behavior
$(i\la_R)^{-3}\, \log(N)^{-2}$.
$Z$ is then determined through
\be
Z = r\,\frac{i\la }{i\la_R}  \quad \sim  \log(N)^{-2}
\label{Z-flow}
\ee
since $r = \frac 1{x^2 Z}$ is finite, and similarly
\be
x = \frac 1{\sqrt{r Z}} \quad  \sim \log(N).
\label{x-flow}
\ee
This gives $a_2$ using \eq{x-def}. 
$a_1$ is determined from \eq{c-def},
\be
Z^2 \, a_1 = \frac{c}{i\la_R} - f_0 
\label{a1-flow}
\ee
whose leading term is linearly divergent in $N$, and
gives the mass renormalization
\be
\delta \mu^2 \theta  
= - \frac{2(i\lambda_R)^2}{16\pi^3 \theta^2}\, \frac 1r Z^2 a_1 \,
\approx  \frac{2(i\la_R)^2}{ 16\pi^3\theta^2 }\, \frac{f_0}{r} 
\label{delta-mu-planar}
\ee
using \eq{delta-mu} up to finite corrections. Its
leading dependence on $N$ is given by \eq{f0-scaling}.
Finally,  $d$ is determined by the constraint
\eq{b-def},
\be
d = 2 b 
+ 2 (i\la_R)^2 xZ \sum_{i}\frac 1{\sqrt{\tilde J_i^2 + 2  b }} 
\label{d-eq}
\ee
which involves the further finite parameter $ b $.
Then $a_0$ follows from \eq{d-def},
which is quadratically divergent in $N$ in agreement with the
perturbative result \eq{a0-pert}.

Finally, \eq{F-derivative-4} shows that $\a=0$ marks
some singularity or phase transition,
dividing the moduli space into two disconnected components with $\a
\gtrless 0$. The explicit form of the 2-point function 
\eq{2point-exact} below shows that the phase with $\a<0$ is the
``physically relevant'' one, while  
$\langle\phi_{ij}\phi_{ji}\rangle <0$ for $\a>0$ for small indices
$i,j$. Since the matrices are supposed to be Hermitian, this
 signals an instability or 
condensation for the low modes, i.e. some kind of tachyonic behavior. 
Observe also that e.g. $\langle \phi_{00}\rangle =0$
can only be realized for $\a<0$.
Nevertheless, the renormalized $n$-point functions remain well-defined
for $\a>0$.
This phase transition is only seen in the genus 0 sector.

Finally consider the case $ b  = 0$, so that
$y_k \equiv \tilde J_k$. Then finiteness of 
the one-point function  requires only that the three quantities
\be
(i\la_R,\, c, \,\frac{1}{i\la_R \a}+ \frac{r}{i\la_R}) 
\label{onepoint-finite-bzero}
\ee
are finite. However, the genus zero 
result for the 2-point function \eq{2point-exact} shows that 
$\frac{1}{\a}$ must also be finite, and we are back to
the previous analysis with finite $\a \neq 0$.

Putting these results together and recalling
the  structure of the higher genus contributions $F_g$ 
stated below \eq{F-higher},
we have established the following:

\begin{itemize}
\item[]{\bf Theorem 1} 
{\em
All derivatives of $F_g$  w.r.t. 
$\tilde J_k^2$ for $g \geq 0$ 
(as well as all functions $F_g$ for $g \geq 2$)
are finite and have a well-defined limit $N \to \infty$, 
provided the 6 quantities $(i\la_R,\a, \mu^2_R, r,c, b )$
are finite and fixed, and $\a \neq 0$. }
\end{itemize}   
This determines the scaling of the 6 bare parameters
\be
(i\la(N),Z(N),\mu^2(N),a_0(N),a_1(N),a_2(N))
\ee
 as a function of $N$
through  \eq{la-flow},  \eq{Z-flow},  \eq{delta-mu-planar},
 \eq{d-eq} resp. \eq{d-def}, \eq{a1-flow}, and \eq{x-flow}.
This defines a renormalization group flow in the space of free parameters.
If desired, we could now impose some specific renormalization 
conditions such as $\langle\phi_{00}\rangle = 0$ etc. These
would then determine the renormalized parameters 
$i\la_R$ etc., which in turn would determine the 
bare ones.

The most interesting ``real'' sector is given by
$i \la_R \in \R, \a< 0, r> 0,\mu_R^2 >-3$ and $b \in \R$ such that 
$\tilde J_k^2 + 2  b >0$ for all $k$. Then $i \la \in \R$
for large enough $N$, and $Z$ and $x$ are positive.
The case $\a=0$ is a singularity of the genus 0 sector.

Since the connected $n$-point functions are given by the derivatives
of $F = \sum_{g \geq 0} F_g$ w.r.t. $\tilde J$, this implies that
all contributions in a genus expansion of the
correlation functions for diagonal entries 
$\langle \phi_{kk} ... \phi_{ll}\rangle$ are finite and well-defined. 
The general non-diagonal correlation functions are
discussed in section \ref{sec:general-correl}, 
and also turn out to be finite 
for arbitrary genus $g$ under the same conditions.
Putting these results together we have
established renormalizability of the model
to all orders in a genus expansion, i.e.
\begin{itemize}
\item[]{\bf Theorem 2} 
{\em All connected genus $g$ contributions 
to any given $n$-point function 
$\langle \phi_{i_1 j_1} ... \phi_{i_n j_n}\rangle_c$
are finite and have a 
well-defined limit $N \to \infty$ for all $g$, under the above conditions.}
\end{itemize}   
Moreover, they can in principle 
be computed explicitly using the above formulas.
This immediately extends to non-connected diagrams 
$\langle \phi_{i_1 j_1} ... \phi_{i_n j_n}\rangle$.
Furthermore, \eq{genus-expansion-power} shows 
that any contribution to 
$F_g$  has order at least $((i\la_R)^2\a)^{2g-1}$.
This implies (but is stronger than) perturbative renormalizability 
to all orders in $\la_R$. This might not seem
surprising in view of the results in
\cite{Grosse:2003nw,Grosse:2004yu,Grosse:2005da}, however note that 
the present model is more complicated than the $\phi^4$ model where
the beta function was found to be zero at one loop 
for $\Omega=1$ \cite{Grosse:2004by}.

It is worth pointing out that only the genus 0 contribution 
requires renormalization, while all higher genus contributions are then
automatically finite. This is very interesting because 
the genus 0 contribution
can be obtained by various techniques in more general models,
see e.g. \cite{Langmann:2003if}.
A  related approach was studied in 
\cite{Steinacker:2005wj,Steinacker:2005tf} without the oscillator
potential.

It may appear surprising to find a well-defined $\phi^3$
model for real coupling, where the action is not bounded from below. 
This is possible  because we {\em first} quantize  
the model for imaginary coupling, where the Kontsevich model is well-defined. 
The full genus expansion is then available, and the limit $N \to \infty$
is under control upon proper renormalization.
This  allows in a {\em second} step to define the model for real coupling 
i.e. real $i\la_R$, through analytic continuation. 
Similar behavior is well-known in the context of 
pure matrix models, which typically allow extension to
seemingly unstable potentials \cite{Brezin:1977sv}; this is 
usually interpreted as suppression of tunneling from a local minimum. 

\subsubsection{Exact renormalization and fine-tuning}

The scaling \eq{la-flow} of the bare coupling is derived assuming that the 
one-point function \eq{F-derivative-4} and therefore its coefficients 
$\a$ and $\la_R$ are kept fixed and independent of $N$. This is the usual
way to proceed in perturbative renormalization, and it would lead to 
new infinite renormalization at each order of perturbation theory, 
as can be seen by
expanding \eq{la-flow} formally in  terms of $\la_R$.

On the other hand, we have the full renormalization in closed form available,
which shows a simple {\em leading} scaling law for large $N$. 
It is then natural to ask what would happen if we scale 
the bare coupling 
in a simpler way, e.g. respecting only the leading scaling 
\be
i \la = c_\la\, \log(N)^{-2}, \qquad
Z = c_Z \,\log(N)^{-2}, \qquad x = c_x \,  \log(N)
\label{approx-scaling}
\ee
for some constant $c_\la, c_Z, c_x$
rather than the exact form \eq{la-flow} etc. It turns out that this is
{\em not} sufficient, and the
renormalization must respect the exact form of \eq{la-flow} ff. rather
than just their leading terms.

To see this, we determine the quantities 
$(\la_R,\a, r, ...)$ for the simpler scaling
\eq{approx-scaling}, and check whether
they also converge as $N \to \infty$. 
This is obviously the case for 
$\la_R \sim \frac{c_\la}{c_x^2 c_Z^2}$ and $r \sim \frac 1{c_x^2
  c_Z}$, while $\a$ is given by \eq{xZ-eq}
\be
 \frac{1}{\a} = - \frac{1}{x Z}  - (i\la_R)^2 f_1 
= -\frac{1}{c_x c_Z} \, \log(N) 
+ \frac 12 \frac{c_\la^2}{c_x^4 c_Z^4}\frac{1}{(16\pi^3 \theta^2)^3} \, \log(N) 
\label{alpha-eq}
\ee
For this to have a well-defined limit $N \to \infty$, 
the above scaling \eq{approx-scaling} is not
sufficient, but the constants of
proportionality must satisfy $\frac{1}{c_x c_Z} 
= \frac 12 \frac{c_\la^2}{c_x^4 c_Z^4}\frac{1}{(16\pi^3
  \theta^2)^3}$, and moreover the scaling must be refined to give
the sub-leading constant \eq{alpha-eq}.

This shows that even this relatively simple, solvable
asymptotically free model has a rather severe ``fine-tuning
problem'', i.e. there is no obvious naturalness in the scaling 
of the bare parameters.

\subsubsection{General $n$-point functions}
\label{sec:general-correl}

Finally we show that all contributions in the genus expansion (and therefore
perturbative expansion) of any $n$-point
functions of the form
\be
\langle \phi_{i_1 j_1} ....  \phi_{i_n j_n}\rangle 
\label{correl}
\ee
have a well-defined and finite limit as $N \to \infty$,
provided the above renormalization conditions hold. 
The argument is the same as in \cite{Grosse:2006qv}, which we 
repeat here for convenience.

Recall that
the insertion of a factor $\tilde\phi_{ij}$ can be obtained by acting
with 
$2i\la_R\frac{\partial}{\partial \tilde J^2_{ij}}$ on $Z(\tilde J^2)\,$, resp. 
$F_g(\tilde J^2)$. 
Now any given correlation function of type \eq{correl}
involves only a finite set of indices $i,j,...$ . Thus taking 
derivatives w.r.t. $\tilde J^2_{ij}$ amounts to considering  
matrices $\tilde J$ of the form 
\be
\tilde J = \left(\begin{array}{ccc} \diag(\tilde J_{1},... \tilde
    J_k) + \delta \tilde J_{k\times k} &\vline & 0 \\
                   \hline 
                  0   & \vline & \diag(\tilde J_{k+1},... \tilde J_N)\end{array}\right),
\label{J-block}
\ee
where $k$ is chosen large enough such that all required variations
are accommodated  by the general hermitian $k\times k$ matrix
\be
\tilde J_{k\times k}:= \left(\diag(\tilde J_{1},... \tilde
  J_k)+\delta \tilde J_{k\times k}\right)
\ee
in \eq{J-block},  while
the higher eigenvalues $\tilde J_{k+1},... \tilde J_N$ are fixed and given by 
\eq{und-J-explicit}. Therefore we can restrict ourselves to 
this $k \times k$ matrix, which is independent of $N$.
As was shown in section \ref{sec:renormaliz}, all $F_g$ 
are in the limit $N \to \infty$ 
smooth (in fact analytic) symmetric functions of the first $k$
eigenvalues squared, hence of the eigenvalues 
of $(\tilde J_{k\times k})^2$. Such a function 
can always be written as a smooth (analytic) function 
of some basis of symmetric polynomials in the $\tilde J_a^2$,
in particular
\be
F_g(\tilde J_1^2, ..., \tilde J_k^2) = f_g(Tr(\tilde J^2_{k\times
  k}), ..., Tr(\tilde J^{2k}_{k\times k})).
\label{F-g-symm}
\ee
This can be seen by approximating the analytic function
$F_g(z_1, .., z_k)$ at the point $z_i = \tilde J_i^2$
by a totally symmetric polynomial in the $z_i$,
which correctly reproduces the partial derivatives up to some order $n$. 
According to a well-known theorem, 
that polynomial can be rewritten as polynomial in the 
elementary symmetric polynomials, or equivalently 
as a polynomial in the variables 
$s_n:=\sum z_i^n$, $n=1,2, ..., k$. This amounts to the
rhs of \eq{F-g-symm}.

In the form \eq{F-g-symm}, it is obvious that all partial derivatives
$\frac{\partial}{\partial \tilde J^2_{ij}}$ of $F_g$ exist to any given order,
and could be worked out in principle.
This completes the proof that each genus $g$ contribution to 
the general (connected) 
correlators $\langle \phi_{i_1 j_1} ....  \phi_{i_n j_n}\rangle $
is finite and convergent as $N \to \infty$.

\subsection{Asymptotic behavior of the correlation functions}
\label{sec:asymptotic}

\subsubsection{$\langle\phi_{kk}\rangle$.}

Using \eq{fR-asymptotic}, the behavior of $\langle\phi_{kk}\rangle$
for large $k$ is dominated by 
$f_R(y_k)$, so that 
\be
\langle \phi_{kk}\rangle_{g=0} \sim \frac 12 
 \frac{i\la_R}{(16\pi^3 \theta^2)^3} \,y_k \log(\frac {y_k}{\tilde J_0})
\qquad\mbox{for}\,\, k \to \infty.
\label{onepoint-asympt}
\ee

\subsubsection{$\langle\phi_{kl}\phi_{lk}\rangle$.}

We can use \eq{2point-eom} to obtain the genus 0 contribution 
to the 2-point function
$\langle\phi_{kl}\phi_{lk}\rangle$ for $k \neq l$. 
Using \eq{F-derivative-4} we obtain the exact expression
\bea
\langle\phi_{kl}\phi_{lk}\rangle_{g=0}
&=& - \frac{2\a^{-1}}{y_k + y_l} + 2(i\la_R)^2
\frac{f_R(y_k)- f_R(y_l)}{y_k^2- y_l^2}
\label{2point-exact}
\eea
Note that the free case corresponds to
$i\la_R =0$ and $\a = -1$. 
Then indeed also the higher genus contributions vanish.

Consider the behavior for large $k \approx l$, which for large indices
is dominated by the terms involving $f_R$:
\bea
\langle\phi_{kl}\phi_{lk}\rangle 
&\approx& 2 (i\la_R)^2
\frac{f_R(y_k)- f_R(y_{l})}{y_k^2 - y_{l}^2}\nn\\
&\approx& (i\la_R)^2 \frac{1}{y_k}\frac{d}{d y_k} f_R(y_k) \nn\\
&\approx& (i\la_R)^2
\frac{1}{2(16\pi^3 \theta^2)^3}\,\frac{\log (y_k/J_0)}{y_k}
 \sim \frac{\log (k)}{k } 
\label{prop-asymptot}
\eea
for large $k\approx l$.
It is worth pointing out that this 2-point function (and similar
the 3-point function \eq{3point-asympt} below, etc.) 
is essentially determined by a function of a single variable $y$,
which describes the dependence of the one-point
function $\langle\phi_{kk}\rangle$ on the 
index $k$. 
This is characteristic for the genus 0 sector, which is
essentially determined by an eigenvalue distribution.

\subsubsection{$\langle\phi_{ll}\phi_{kk}\rangle$.}

As a further example, consider the 2-point function 
$\langle\phi_{ll}\phi_{kk}\rangle$ for $k \neq l$, 
which vanishes in the free case.
To compute it from the effective action, we need in principle
\bea
\langle\phi_{ll} \phi_{kk}\rangle_c
= \langle\tilde\phi_{ll}\tilde\phi_{kk}\rangle - \langle\tilde\phi_{kk}\rangle
\langle\tilde\phi_{ll}\rangle 
&=& (2i\la_R)^2\frac{\partial}{\partial \tilde J_l^2}
\frac{\partial}{\partial \tilde J_k^2}(F_0 + F_1 + ...) .
\eea
Even though this corresponds to a nonplanar diagram with external legs, 
it is  obtained by
taking derivatives of a closed genus 0 ring diagram.
Therefore we expect that only $F_0$ will contribute, 
and indeed the derivatives of $F_1$ contribute only to order
$\la_R^{4}$.
For $k \neq l$, the genus zero contribution is
\bea
\langle\phi_{ll} \phi_{kk}\rangle_c &=& 
(2i\la_R)^2\frac{\partial}{\partial \tilde J_l^2}
\frac{\partial}{\partial \tilde J_k^2} F_0 \nn\\
&=& (i\la_R)^2
\frac{1}{\sqrt{\tilde J_k^2+2 b }}\frac{1}{\sqrt{\tilde J_l^2+2 b }}
\left(\frac{1}{\sqrt{\tilde J_k^2+2 b } +\sqrt{\tilde J_l^2+2 b }}\right)^2  
\nn\\
&\approx& (i\la_R)^2
\left(\frac{1}{\tilde J_k^2+2 b }\right)^2
\label{prop-kk-asymptot}
\eea
as  $k \approx l \to \infty$,
using again \eq{constr-implicit-2}.
For $k=l$, we need
\be
\langle\phi_{kk} \phi_{kk}\rangle_c = 
(2i\la_R)^2\frac{\partial^2}{\partial (\tilde J_k^2)^2} F_0 
\approx \frac{(2i\la_R)^2}2 \, 
 \frac{\partial}{\partial \tilde J_k^2} f_R(y;\delta \tilde J),
\label{2point-kk}
\ee
since $f_R(y)$ is 
the leading contribution for large $k$. 
Writing
\bea
\frac{d}{d \tilde J_k^2} f_R(y;\delta \tilde J) &=& 
\frac{d y }{d \tilde J_k^2}\frac{\partial}{\partial  y} 
f_R(y;\delta \tilde J)
+ \frac{\partial}{\partial \tilde J_k^2}\(y_k^2 
\frac 1{(\tilde J_k^2+2 b )\Big(y_k \sqrt{\tilde J_k^2+2 b }
   + (\tilde J_k^2+2 b )\Big)} \),  \nn\\
\label{2point-coincident}
\eea
we note that the first term (involving the sum) is dominant,
while the second is only one term in a large sum.
This then gives the same asymptotic behavior as the 
usual propagator,
\be
\langle\phi_{kk} \phi_{kk}\rangle_c 
\approx \frac{(2i\la_R)^2}2 \, 
\frac{\partial}{\partial y_k^2} f_R(y_k;\delta \tilde J)
\approx \langle\phi_{kl}\phi_{lk}\rangle 
\label{prop-asymptot-2}
\ee
for large $k$, using   \eq{prop-asymptot}.
On the other hand, it is very different from 
$\langle\phi_{ll} \phi_{kk}\rangle_c$ given by \eq{prop-kk-asymptot}.
This should not be a surprise, since the latter 
corresponds to nonplanar diagrams as in figure \ref{fig:nonplanar}.

Note that in \eq{2point-coincident},  
a derivative 
w.r.t. $\tilde J_k^2$ has essentially been replaced by  a derivative
w.r.t. $y\equiv y_k$. 
This is again characteristic for the genus 0 sector. 
In fact, $F_0$ can be obtained from the Dyson-Schwinger equation by making 
precisely this approximation.

\subsection{The 3-point function}

We can similarly 
use the exact expression \eq{3point-2} for the 3-point function to 
determine the asymptotic behavior of the 3-point function  at genus 0:
\bea
\left<\phi_{kl}\phi_{lk} \phi_{kk}\right>_{c} 
&=& 2i\la_R \frac{1}{\tilde J_k^2-\tilde J_l^2}\,
\(\langle(\phi_{kk} -  \phi_{ll}) \phi_{kk}\rangle_c
- \langle\phi_{kl}\phi_{lk}\rangle\) \nn\\
&\approx& 2i\la_R \frac{1}{\tilde J_k^2-\tilde J_l^2}\,
\(\langle \phi_{kk} \phi_{kk}\rangle_c
- \langle \phi_{kl} \phi_{lk}\rangle\) \nn\\
&\approx& (2i\la_R)^3 \frac{1}{y_k^2-y_l^2}\,
\frac 12\( \frac{\partial}{\partial y_k^2} f_R(y;\delta \tilde J)
- \frac{f_R(y_k)- f_R(y_{l})}{y_k^2 - y_{l}^2}\rangle\) \nn\\
&\approx& (i\la_R)^3 2
( \frac 1{2y_k} \frac{d}{d y_k} )^2 f_R(y_k;\delta \tilde J)\nn\\
&\approx& -(i\la_R)^3 \frac 14 \frac{1}{(16\pi^3 \theta^2)^3}\,
\frac 1{y_k^3} \log(\frac {y_k}{\tilde J_0})
\label{3point-asympt}
\eea
for large $k \approx l$,
using \eq{prop-kk-asymptot}, \eq{prop-asymptot} and \eq{fR-asymptotic}.
The same behavior is found for
\be
\langle\phi_{kk} \phi_{kk} \phi_{kk}\rangle_c =
(2i\la_R)^3\frac{\partial^3}{\partial (\tilde J_k^2)^3} F_0 \nn\\
\approx (2i\la_R)^3 \frac 12
\frac{\partial^2}{\partial (y_k^2)^2} f_R(y_k),
\ee
up to a factor 2 which reflects the exchange symmetry between the
external ``legs'' $\phi_{kl}$ and $\phi_{lk}$ for $k=l$.
Combining \eq{3point-asympt} with \eq{prop-asymptot} and 
\eq{prop-asymptot-2},
the 1PI vertex behaves like
\bea
\left<\phi_{kl}\phi_{lk} \phi_{kk}\right>_{1PI}
&=& \frac 1{\left<\phi_{kl} \phi_{lk} \right>_c} 
\frac 1{\left< \phi_{kl} \phi_{lk}\right>_c} 
\frac 1{\left<\phi_{kk} \phi_{kk} \right>_c} 
\left<\phi_{kl}\phi_{lk} \phi_{kk}\right>_{c} \nn\\
&\approx& -2(i\la_R)^{-3}(16\pi^3 \theta^2)^6\,
\(\frac{1}{\log (y_k/\tilde J_0)}\)^2 
\sim  -\(\frac{1}{\log k}\)^2 
\label{3point-1PI-asympt}
\eea
for large $k \approx l$. This result is exact, because the higher genus 
contributions decay more rapidly as discussed in the next section. 
In particular, this establishes asymptotic freedom.

Several remarks are in order. 
First we note 
that this vertex function decays like $\frac 1{(\log k)^2}$ rather
than $\frac 1{(\log k)^{1/2}}$, which would be found by a 1-loop computation
as shown in section \ref{sec:perturbative}.
Moreover, \eq{3point-1PI-asympt} is in nice agreement with
\eq{la-flow}, demonstrating that
the 1PI vertex approaches the bare coupling $-\frac{i\la}2$ 
for $k \to N$.
Finally, note that the effective coupling constant for this
asymptotic domain is $\frac 1{\la_R^3}$ rather than $\la_R$. 
This is clearly a purely ``quantum'' effect,
which is again due to higher-order correction 
to the one-loop RG result \eq{g-running-1loop}.

\subsection{Higher genus contributions}

We illustrate here the stronger 
decay behavior of the higher-genus $n$-point functions, 
in the example of the one-point function. 
Consider
\bea
\langle\phi_{kk}\rangle_{g=1}
&=& 2i\la_R \frac{\partial}{\partial \tilde J_l^2}\, F_1 
= -\frac{2i\la_R}{24} \frac{\partial}{\partial \tilde J_k^2}\, 
\ln(1-I_1) \nn\\
&=& \frac{(i\la_R)^3\a }{12} 
\,\frac{\partial}{\partial \tilde J_k^2}\, f_{1,R}(\d\tilde J)   \nn\\
&=& \frac{ (i\la_R)^3 \a}{8} 
\,\frac1{(\tilde J_k^2+2 b )^{5/2}}
(1+2\frac{\partial}{\partial \tilde J_k^2}\, b )  \nn\\
&=& \frac{(i\la_R)^3 \a}{8} 
\,\frac1{(\tilde J_k^2+2 b )^{5/2}}
\(1-\,\frac{(i\la_R)^2\a}{\sqrt{\tilde J_k^2 + 2 b }^3}\) 
\label{genus-1}
\eea
using \eq{1-I-reg-2} and \eq{b-derivative}.
This asymptotically behaves like 
\be
\langle\phi_{kk}\rangle_{g=1} \sim \frac1{k^{5}}
\label{higher-genus-decay}
\ee
for $k \to \infty$. 
Looking at \eq{F-higher}, it is obvious that the higher-genus
contributions are also decaying at least as rapidly as 
the genus 1 contribution \eq{higher-genus-decay}, and similar results 
could be derived for the $n$-point functions.

Finally, it follows from \eq{IkI1} 
 that each term  in $F_g$ contains the factor
\be
F_g \sim ((i\la_R)^2\a)^{2(g-1) + \sum l_p}
\label{genus-expansion-power}
\ee
for $g \geq 2$, which is at least $\a^{2(g-1) + 1}$. Hence $(i\la_R)^2\a$ is 
the  parameter which controls the genus expansion.

\section{Further aspects}
\label{sec:further}

\subsection*{Structural remarks}

We briefly add some heuristic remarks which might 
shed new light on the formal results obtained above.

The model considered here is characterized by a function $F$
of $\tilde J$ which is invariant under conjugation with $U(N^3)$.
This implies in particular that $F = F_N(\tilde J_k)$ is a 
totally symmetric function of the eigenvalues $\tilde J_k$,
i.e. it is a  function on the 
quotient space $\R^{N^3}/\cS_{N^3}$
where $\cS_{N^3}$ denotes the permutation group.
Since the observables are obtained by taking partial derivatives 
w.r.t. $\tilde J$,
we are interested in
\bea
W_k(\tilde J) &=& \frac{\partial}{\partial \tilde J_k} F_N(\tilde J), \nn\\
W_{k,l}(\tilde J) &=& \frac{\partial}{\partial \tilde J_k}
\frac{\partial}{\partial \tilde J_l} F_N(\tilde J), 
\eea
etc. Renormalizability requires that all these 
derivatives exist and have a well-defined limit $N\to\infty$, hence that 
$F_N(\tilde J_i)$ converges to an infinitely differentiable 
function $F(\tilde J_i)$. This must hold at the point
$\tilde J$ with eigenvalues given by \eq{und-J-explicit}, or
preferably in some ``good'' neighborhood $\cU \ni \tilde J$ 
in $\R^{N^3}$, with an appropriate limit $N \to \infty$.
Furthermore, these partial derivatives should have suitable decay properties
such as indicated above. 

Now consider the dependence of $W_{k}(\tilde J)$ on the index $k$.
It is a physical requirement that this dependence on
the indices is mild\footnote{some mild singularities for coinciding indices 
might be allowed however}. This can be understood analytically as follows.
Note that  $W_{k}(\tilde J)$ is related to $W_{l}(\tilde J)$ by 
exchanging the arguments $\tilde J_k$ and $\tilde J_l$,
i.e. by applying the permutation operator $\sigma_{k,l}$ on the space
$\R^{N^3}$ of $\tilde J_k$,
\be
W_{k} =  W_{l} \circ \sigma_{k,l}.
\ee 
This means that the index dependence may be traded for a
``small'' change (a permutation) of the eigenvalues.
Since the eigenvalues are ordered and approach a simple distribution,
it is plausible that $\sigma_{k,l}$ respects the 
neighborhood $\,\cU$. Smoothness in $\cU$  would then naturally imply 
that the dependence on the indices is mild. 
Indeed, we found explicitly that the index dependence at genus 0 becomes 
translated into the dependence of a smooth function $W(y)$ \eq{F-derivative-2}
on the variable $y$.

\subsection*{Extension to $\Omega \neq 1$}
\label{sec:beyondomega}

Once a suitable domain $\cU$ is established where $F[\tilde J]$ is
smooth with suitable decay properties of its partial derivatives, 
then the following strategy to extend our results to $\Omega \neq 1$
can be envisaged. 
The partition function for 
 $0 \neq \Omega \neq 1$ can be obtained from the results for $\Omega=1$ using
\bea
Z[\tilde J] &=& \langle e^{\epsilon Tr[\tilde x_i,\phi][\tilde x_i,\phi]}
\rangle_{\Omega =1}
= e^{\epsilon Tr[\tilde x_i,\frac{\partial}{\partial \tilde J^2}]
[\tilde x_i,\frac{\partial}{\partial \tilde J^2}]} Z[\tilde J]_{\Omega =1} \nn\\
&=& e^{\epsilon Tr[\tilde x_i,\frac{\partial}{\partial \tilde J^2}]
[\tilde x_i,\frac{\partial}{\partial \tilde J^2}]}\, 
e^{F[\tilde J]_{\Omega =1}}
=: e^{F[\tilde J]}
\eea
for small $\epsilon$, accompanied by a change of wavefunction normalization.
Hence we should consider the operator
\be
\Delta_J := Tr[\tilde x_i,\frac{\partial}{\partial \tilde J^2}]
[\tilde x_i,\frac{\partial}{\partial \tilde J^2}]
\ee
acting on functions of $\tilde J$.
It does not commute with $U(N^3)$, hence $Z[\tilde J]$ will be 
a function of $\tilde J$ which  no longer 
depends on the eigenvalues only. We have established in this paper that
$Z[\tilde J]_{\Omega =1}$ is a
well-defined, infinitely
differentiable function 
(after subtracting a possible infinite constant from $F$). 
The question of renormalizability for
$\Omega \neq 1$  is whether $F[\tilde J]$ is also
a well-defined, infinitely
differentiable function, possibly after further renormalization. 
To establish this 
using the methods of the present paper might therefore be feasible,
by studying the operator $\Delta_J$ and 
establishing careful estimates on the partial derivatives of $F[\tilde
J]_{\Omega =1}$ and their asymptotic behavior for large $N$.  However,
we will not attempt to do this in the present paper.

\subsection*{Remarks on the relation with string theory}

The Kontsevich model has been related to  string theory in 
\cite{Gaiotto:2003yb} as follows. 
The matrix $X_{ij}$ of the Kontsevich model is interpreted as
coefficient
of the open string field, more precisely the tachyon, connecting 
the (Liouville) $D$-brane with label $i$ to the $D$-brane with label $j$. 
The eigenvalues $\tilde M_i$ of the external potential in the
Kontsevich model are interpreted as boundary cosmological constants on the
brane with label $i$, and the Kontsevich model \eq{Z-Konts}
describes the 
(topological sector of) open string field theory in this situation.

Applying this interpretation to our model, we are led to a picture
of  open string tachyons in string
field theory with $N^3$ D-branes, with specific cosmological constant
$J_i \sim i_1+i_2+i_3$  on the brane $i = (i_1,i_2,i_3)$.
The fact that our model is renormalizable implies 
that the correlators become essentially 
smooth functions of these 3 coordinates $i_1,i_2,i_3$, 
with a well-defined large $N$ limit.
This could be interpreted by saying that 
the endpoints $i,j$ of the open strings effectively live in
3 dimensions with coordinates 
$i_1, i_2, i_3 \geq 0$, forming a 3-dimensional wedge $\R^3_+$
with a potential determined by $J_i$. 
In other words, 
3 extra dimensions appear from the string point of view
by some kind of stringy ``deconstruction'' of dimensions. 
On the other hand, we  have the interpretation as 6-dimensional
non-commutative field theory. This points to interesting
directions for further  studies.

\section{Perturbative computations}
\label{sec:perturbative}

We write the bare action \eq{action-kontsevich} as
\bea
S &=& Tr \Big(\frac 14 (J \phi^2 + \phi^2 J)
+ \frac{i\la}{3!}\;\phi^3 - (i\la) A \phi \Big) \nn\\
&=&  Tr\Big( \frac 12 \phi^i_j \; (G_R)^{j;l}_{i;k}\;  \phi^k_l 
+ \frac{i\la_R}{3!}\;\phi^3 \Big) + \delta S\, .
\eea
The finite  kinetic term 
$ (G_R)^{j;l}_{i;k}  = \frac 12 \delta^i_l \delta^k_j (\tilde J_i+\tilde J_j)$ 
defines the renormalized propagator
\be
\Delta^{i; k}_{j;l} = \langle \phi^i_j \phi^k_l\rangle 
 = \delta^i_l \delta^k_j \frac 2{\tilde J_i+\tilde J_j}
= \delta^i_l\delta^k_j\frac {1/(4\pi^2 \theta)}{\und{i} +\und{j} +
  (\mu_R^2\theta+3)},
\label{propagator}
\ee
corresponding to the finite (renormalized) matrix
\be
\tilde J |n_1,n_2,n_3\rangle 
= 16\pi^3 \theta^2\Big(\und{n}+\frac{3+\mu_R^2\theta}2\Big)|n_1,n_2,n_3\rangle,
\ee
using the notation 
$\und{n} = n_1+n_2+n_3$ \eq{n-notation}.
Here $\la_R$ is the renormalized coupling constant, 
valid at a  ``scale'' given by $m$, i.e. assuming that
the  indices of the interaction vertex 
approximately satisfy
\be
\und i\approx \und j\approx \und k = \und m.
\label{i-approx}
\ee
The counterterms are collected in
\be
\delta S = Tr \Big(  - (i\la_R) Z_\la A \phi
+ \frac 14 (\d J \phi^2 + \phi^2 \d J)  
+ \frac{i\delta\la}{3!}\;\phi^3  \Big),
\ee
where 
\bea
\d J |n_1,n_2,n_3\rangle 
&=& 16\pi^3 \theta^2\Big((Z\frac{\d\mu^2\theta}2)  + (Z-1) J_R\Big)
|n_1,n_2,n_3\rangle,  \nn\\
\la_R &=& Z_\la^{-1} \la, \nn\\
\delta \la &=& \la_R(Z_\la-1)
\label{deltaJ-def}
\eea
is part of the counter-term, and
$\d\mu^2 = (\mu^2 - \mu_R^2)$.
It is then easy to see that the usual power counting rules apply
(with suitable extensions to the case of higher genus \cite{Grosse:2003aj}),
where $N$ plays the role of the cutoff $\Lambda^2$. This 
can be seen in the following 1-loop examples:

\paragraph{3-point function}

\begin{figure}[htpb]
\begin{center}
\epsfxsize=1.3in
   \epsfbox{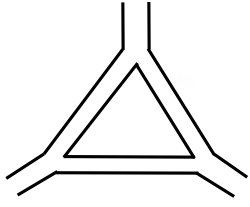}
\end{center}
 \caption{one-loop contribution to 1PI vertex}
\label{fig:planarvertex}
\end{figure}

In 6 dimensions, 
the interaction vertex requires renormalization, induced by
the planar one-loop 1PI graph (without external legs) in figure
\ref{fig:planarvertex}.
This gives 
\bea
\langle \phi_{ij} \phi_{jk} \phi_{ki} \rangle_{1PI}
&=& -\frac{i\la_R}2 + (-\frac{i\la_R}2)^3 \sum_{\und l}
\frac{2}{J_k^R + J_l^R}  \frac{2}{J_i^R + J_l^R} \frac{2}{J_j^R + J_l^R}
 -\frac{i\delta\la}2 \nn\\
&\approx& -\frac{i\la_R}2 + (-i\la_R)^3 \sum_{\und l}
\frac{1}{(J_l^R + J_m^R)^3}  -\frac{i\delta\la}2 \, + finite \nn\\
&\approx&  -\frac{i\la_R}2 + \frac{(-i\la_R)^3}{(16\pi^3 \theta^2)^3} 
\sum_{\und l} \frac{1}{(\und l+\und m)^3}  -\frac{i\delta\la}2 \,\, + finite  \nn\\
&\approx&  -\frac{i\la_R}2 
+  \frac{(-i\la_R)^3}{(16\pi^3 \theta^2)^3}
\frac 12 \log\frac{N+m}{8m} \, -\frac{i\delta\la}2 \, + finite  
\label{vertex1PI-m}
\eea
All nonplanar contributions are finite. This gives
the counterterm to the coupling constant 
\be
i\delta\la =  -\frac{(i\la_R)^3}{(16\pi^3 \theta^2)^3}
 \log\frac{N+m}{8m} \, + finite 
\ee
so that
\be
(i\la) = (i\la_R)
(1 +  \frac{(i\delta\la)}{(i\la_R)})
= \frac{(i\la_R)}{1+\frac{(i\la_R)^2}{(16\pi^3 \theta^2)^3}\, 
\log \frac Nm} \quad + O(\la_R^5) .
\label{renorm-coupl}
\ee
The minus sign indicates asymptotic freedom.
However, the last formula \eq{renorm-coupl} is only suggestive, and
such a one-loop result should be used with caution.
Note that we could equally well write
\be
(i\la) 
= \frac{(i\la_R)}{\(1+\frac 12\frac{(i\la_R)^2}
{(16\pi^3 \theta^2)^3}\, \log \frac Nm\)^2} \quad + O(\la_R^5),
\ee
which is in fact agrees much better with the exact
scaling \eq{la-flow} of the bare coupling.
A better way to understand the coupling constant
 renormalization is to determine
its dependence on the ``scale'' $m$, i.e. the running coupling
constant
\be
g(m) = - i\la_R(m) =  2\langle \phi_{ij} \phi_{jk} \phi_{ki}
  \rangle_{1PI}
\ee
for $\und i\approx \und j\approx \und k = \und m$ \eq{i-approx}.
An extra factor $2$ is inserted for convenience.
The corresponding one-loop beta function can be 
obtained from \eq{vertex1PI-m},
\be
\beta := 2 m\,\frac{\partial\langle \phi_{ij} \phi_{jk} \phi_{ki}
  \rangle_{1PI}}{\partial m} 
= \frac{g(m)^3}{(16\pi^3 \theta^2)^3} 
  \,m \frac{\partial\log\frac{N}{m}}{\partial m }
= - \,\frac{g(m)^3}{(16\pi^3 \theta^2)^3} 
\label{beta}
\ee
indicating asymptotic freedom. This gives
\be
\frac{d g(m)}{g(m)^3} = - \,\frac{1}{(16\pi^3 \theta^2)^3}  \frac {dm} m
\ee
which can be  integrated to give the running (1PI) coupling constant
\be
g(m)^2 = \frac{g(m_0)^2}{1+\frac{2g(m_0)^2}{(16\pi^3 \theta^2)^3}
  \log(\frac m{m_0})}.
\label{g-running-1loop}
\ee
This decreases with increasing scale $m$, 
which means asymptotic freedom.
However, note that the exact scale dependence 
$g(m) \sim  -\frac{1}{(\log m)^2} $
for large $m$ which was determined in 
\eq{3point-1PI-asympt}
is not correctly reproduced by this one-loop RG computation.
This shows that the common practice of using the 1-loop RG
results for the running coupling constant may not be sufficiently
accurate for large scales. 

It is worth pointing out that the coupling constant runs 
here even at one-loop,
in contrast to the case of the $\phi^4$ model in 4 dimensions 
\cite{Grosse:2004by}.

\paragraph{1-point function}
The one-loop contribution to the 1-point function gives
\bea
\langle \phi_{ii} \rangle_{1-loop}
&=&  \frac{i\la_R Z_\la}{\tilde J_i} A_i - \frac{i\la_R}2\,  
\frac 1{\tilde J_i} \sum_{k}  \frac 2{\tilde J_i+\tilde J_k} \quad   \nn\\
&=& -\frac{i\la_R}{\tilde J_i}\left(-  Z_\la A_i 
+ \frac 1{16\pi^3\theta^2} \sum_{k} 
 \frac 1{\und{i} +\und{k} + 3+\mu_R^2\theta}\right).
\label{onepoint-oneloop}
\eea
To proceed, we expand
\be
h(\und i) := \sum_{k} \frac 1{\und{i} +\und{k} +3+ \mu_R^2\theta}
 = h(0) + (\und{i})\, h'(0) + \frac 12(\und{i})^2\, h''(0) 
+  h_R(\und i)
\ee
where $h_R(\und i)$ 
is a finite nontrivial function of $\und{i}$, and
\bea
h(0) &=& \sum_{k}\frac 1{\und{k} + \mu_R^2\theta+3} 
=  (-6\log 2 + \frac 92 \log 3)N^2 \nn\\
&& +  (-6 (\mu_R^2\theta+3) \log 2 + 3(\mu_R^2\theta +3) \log 3) N 
 +  \frac 12 (\mu_R^2\theta +3)^2\log N ,\nn\\ 
h'(0) &=& - \sum_{k} \frac 1{(\und{k} + \mu_R^2\theta+3)^2} 
  = -(16\pi^3\theta^2)^2\, f_0 - (16\pi^3\theta^2)^3
  (3+\mu_R^2\theta)\,f_1  \nn\\
&=& - (6\log(2)-3\log(3))N +  \frac{3+\mu^2_R\theta}2 \log(N) , \label{hprime} \\
h''(0) &=& 2\, \sum_{k} \frac 1{(\und{k} +
  \mu_R^2\theta+3)^3} =  - (16\pi^3\theta^2)^3\, 2 f_1 = \log(N)  
\label{hpprime}
\eea
up to finite corrections,
using the results of section \ref{sec:f-renormaliz}.
We note in particular that the $i$- dependent term 
$(\und{i})\, h'(0)+ \frac 12(\und{i})^2\, h''(0) $ in 
\eq{onepoint-oneloop} forces us to introduce a corresponding
counterterm to the action as in \eq{A-def},
$A = a_0 + a_1 J + a_2 J^2$.
As discussed in section \ref{sec:phi3}, 
this is no longer equivalent to an infinite shift \eq{phi-shift}
of $\phi$. 
Taking this into account we have
\bea
\langle \phi_{ii} \rangle_{1-loop}
&=& -\frac{i\la_R}{\tilde J_i}\Big(-(a_0 + a_1 J + a_2 J^2) Z_\la  \nn\\
&& \quad + \frac 1{16\pi^3\theta^2} 
(h(0) + (\und{i})\, h'(0) +  \frac 12(\und{i})^2\, h''(0) 
+  h_R(\und i))\Big).
\label{onepoint-oneloop-2}
\eea
Finiteness and the condition $\langle \phi_{00}\rangle =0$ implies
to lowest order
\bea
a_2 &=&  \frac 1{(16\pi^3\theta^2)^3} \,\frac 12\log N
  = - f_1 + finite, \label{a2-pert}\\ 
a_1  &=& -(16\pi^3\theta^2) (\mu^2_R\theta +3)  a_2
+ \frac 1{(16\pi^3\theta^2)^2} \,h'(0) 
 = - f_0  + finite , \label{a1-pert}\\ 
a_0 &=&\frac 1{16\pi^3\theta^2} \,h(0) 
-  (8\pi^3\theta^2)(\mu^2_R\theta +3)  a_1 
-  (8\pi^3\theta^2)^2(\mu^2_R\theta +3)^2  a_2  
\label{a0-pert}
\eea
up to finite corrections.
These renormalization conditions guarantee that the one-point function
$\langle \phi_{ii} \rangle$ has a well-defined 
and nontrivial limit $N\to\infty$.
In particular, \eq{a2-pert} and \eq{x-def} imply
\be
x = \sqrt{1+ \frac{(i\lambda_R)^2}{(16\pi^3\theta^2)^3} \,\log N}
= 1+\frac 12 \frac{(i\lambda_R)^2}{(16\pi^3\theta^2)^3} \,\log N,
\ee
to lowest order, and \eq{a1-pert} together with \eq{delta-mu} gives 
\be
\d\mu^2\theta =  \frac{2(i\la_R)^2}{(16\pi^3\theta^2)}\, f_0 + finite.
\label{mu-renorm-pert}
\ee
This is consistent with the exact result \eq{delta-mu-planar}.

Finally, it is interesting to consider the behavior of 
$\langle \phi_{ii} \rangle$ for large $i$. After imposing the 
above renormalization conditions, we observe that the dominating term
in \eq{onepoint-oneloop-2} is 
$\frac{i\la_R}{\tilde J_i} \frac 12(\und{i})^2\, h''(0)$.
Here $h''(0) \sim \log(N)$ should more properly be replaced by 
$\log(N/\und i)$, which implies 
\be
\langle \phi_{ii} \rangle \sim \und{i}\, \log(\und i)
\ee
for large $i$. This is in agreement with \eq{onepoint-asympt}.

\begin{figure}[htpb]
\begin{center}
\epsfxsize=2in
   \epsfbox{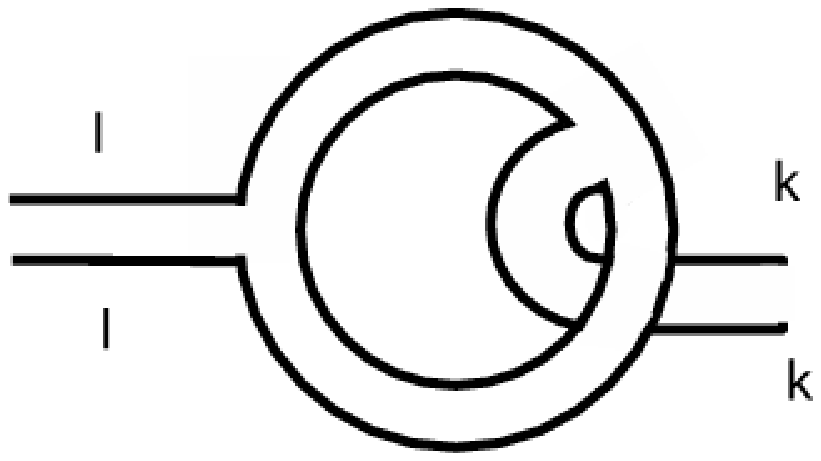}
\end{center}
 \caption{one-loop contribution to $\langle\phi_{ll}\phi_{kk}\rangle $}
\label{fig:nonplanar}
\end{figure}

\paragraph{2-point function}
Next we compute the leading contribution to the 
2-point function  $\langle\phi_{ll}\phi_{kk}\rangle$ for $l\neq k$,
which vanishes at tree level. The leading contribution 
comes from the nonplanar graph in figure \ref{fig:nonplanar},
which gives
\bea
\langle\phi_{ll}\phi_{kk}\rangle 
= \langle\phi_{kk}\rangle \langle\phi_{ll}\rangle 
+\frac 14 \frac{(i\la_R)^2}{\tilde J_k\, \tilde J_l}\left(\frac{2}{\tilde J_k + \tilde J_l}\right)^2  
\label{2point-nonplanar}
\eea
(for $l\neq k$) indicating the symmetry factors, 
where the disconnected contributions 
are given by \eq{onepoint-oneloop}.
This is consistent with the result 
\eq{prop-kk-asymptot} obtained from the 
Kontsevich model approach. Note that the counterterm $\d J$ does not
enter here.

\begin{figure}[htpb]
\begin{center}
\epsfxsize=3.5in
  \vspace{0.2in} 
   \epsfbox{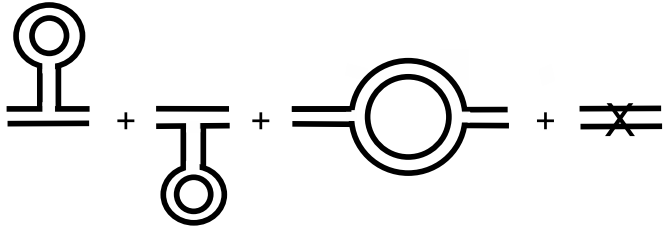}
\end{center}
 \caption{one-loop contribution to $\langle\phi_{kl}\phi_{lk}\rangle $}
\label{fig:planarprop}
\end{figure}

Similarly, the leading contribution to the 
2-point function  $\langle\phi_{kl}\phi_{lk}\rangle$ for $l\neq k$
has the contribution indicated in figure \ref{fig:planarprop}, 
involving also the counterterm $\d J$. 
This gives
\bea 
\langle\phi_{kl}\phi_{lk}\rangle_{1-loop}
&=& \frac{2}{J_k^R + J_l^R} 
- 2(i\la_R) \frac{\langle \phi_{kk} + \phi_{ll}\rangle}{(J_k^R + J_l^R)^2} \nn\\
&&
 + \frac{4}{(J_k^R + J_l^R)^2} \Big(\sum_{j}
  \frac{(i\la_R)^2}{J_k^R +J_j^R}\frac{1}{J_l^R +J_j^R}  
  - \frac{\d  J_l+\d J_k}2 \Big).
\label{2point-eom-2-pert}
\eea 
The first term is the free propagator, 
the next term the tadpole contributions, and
the last them the one-loop contribution in figure \ref{fig:planarprop}
with counterterm $\delta J$.

We  have to adjust the parameters such that the result is
well-defined and nontrivial. Using \eq{hprime} and \eq{hpprime}, we can write
\bea
\sum_{j} \frac{1}{J_k^R +J_j^R}\frac{1}{J_l^R +J_j^R}
&=& \frac 1{(16\pi^3\theta^2)^2}
\sum_{j}\(\frac{1}{(\und{j}+\mu_R^2\theta+3)^2} 
 - \frac{\und k +  \und l}{(\und{j}+\mu_R^2\theta+3)^3} \) + finite \nn\\
&=& \frac 1{(16\pi^3\theta^2)^2} \, 
\Big(-h'(0) - (\und k + \und l) \frac 12 h''(0) + finite\Big) 
\label{prop-divergences}
\eea
Therefore 
\bea
\d J_l+\d J_k &=&  \frac{2 (i\la_R)^2}{(16\pi^3\theta^2)^2} \,
\Big( -h'(0) - (\und k + \und l) \frac 12 h''(0)  \Big)\,\, + finite,
\eea
which using \eq{hprime}, \eq{hpprime} implies
\be
 (Z-1) (\und k + \und l) 
= 2 (i\la_R)^2 \,(\und k + \und l)  f_1 
\ee
and
\be
Z (\delta\mu^2\theta)  + (Z-1) (3+\und \mu_R^2\theta)
=  \frac{2 (i\la_R)^2}{(16\pi^3\theta^2)} f_0 + 2 (i\la_R)^2 (3+\mu_R^2\theta)\,f_1  
\ee
up to finite corrections. Hence we obtain the lowest order mass 
and wavefunction renormalization:
\bea
Z &=& 1+ 2 (i\la_R)^2 \, f_1 
 = \frac{1}{\(1+\frac 12\frac{(i\la_R)^2}{(16\pi^3 \theta^2)^3}\,
   \log N\)^2} \quad + O(\la_R^4)\nn\\
\d\mu^2\theta &=&  \frac{2(i\la_R)^2}{(16\pi^3\theta^2)}\, f_0 
\label{mu-renorm-pert.2}
\eea
up to finite corrections, in agreement with \eq{mu-renorm-pert}.

Note that in 8 or higher dimensions,  divergent mixed terms 
$(\und k)(\und l)$ would occur in \eq{prop-divergences}
which can no longer be absorbed. Then the model 
is no longer renormalizable, as in the commutative case.

\section{Summary and discussion}

We have shown that the selfdual NC
$\phi^3$ model in 6 dimensions can be renormalized and essentially solved
in terms of a genus expansion, by using the Kontsevich model.
This provides a model which contains essentially the full
complexity of  renormalization of a 
not super-renormalizable asymptotically free quantum field theory, while 
being solvable and hence fully under control.
In principle, all $n$-point functions can be computed in a genus
expansion, and we give explicit expressions for the 1-, 2-, and some
3-point functions.

In particular, we were able to determine exactly the RG flow
of the bare parameters as a function of the cutoff $N$, 
as well as the running of the ``physical'' 
coupling constant i.e. the 1PI 3-point function.
As in the case of 2 and 4 dimensions
\cite{Grosse:2005ig,Grosse:2006qv}, it turns out that the
renormalization is fully determined by the genus 0 sector. 
In particular, we can compare the exact results with the  
standard perturbative methods.
For example, it turns out  that the 1-loop beta function
 for the coupling constant gives
roughly the correct behavior and correctly predicts  asymptotic
freedom, but wrongly gives a $(\log N)^{-1}$ dependence on the scale
as opposed to the correct $(\log N)^{-2}$ dependence.

We also show that the model has a
 critical surface defined by $\a=0$, which separates  
2 different phases. One phase has the expected ``physical''
properties, while in the other some modes become unstable.

It is very remarkable that a nontrivial, 
asymptotically free 6-dimensional NC $\phi^3$ field theory 
allows such a detailed analytical description.
There is no commutative analog where this has been achieved
to our knowledge.
Therefore this model can serve as a testing ground for various 
ideas and methods for renormalization.
 It also  shows that the noncommutative world
in some cases is 
more accessible to analytical methods than the commutative case.
While the techniques used in this paper are
more--or--less restricted to the $\phi^3$ interaction, it is worth 
pointing out that the renormalization 
is determined by the genus 0 contribution only, which is
accessible in a wider class of models; see also 
\cite{Steinacker:2005wj,Steinacker:2005tf} in this context.

One of the open problems is the lack of control
over the {\em sum } over all genera $g$. 
We have not shown that the sum over $g$ 
converges in a suitable sense, which would amount 
to a full construction of the model. 
Furthermore, it would be extremely interesting to extend the
analysis beyond the case of $\Omega=1$. 
We propose a strategy in section \ref{sec:further}  how 
such an extension might be possible, and we also comment 
on a relation with open string field theory.

\paragraph{Acknowledgements}

We are grateful for discussions with B. Eynard, I. Kostov,
E. Langmann, J. Magnen, R. Wulkenhaar,  and F. Vignes-Tourneret.
This work was supported by the FWF project  P18657.

\bibliographystyle{diss}

\bibliography{mainbib}

\end{document}